\documentclass[11pt]{article}
\usepackage[top=1in,bottom=1in,left=1in,right=1in]{geometry}
\usepackage{authblk}
\usepackage{blindtext}
\usepackage{amsmath,amssymb,amsthm}
\usepackage[T1]{fontenc}
\usepackage[tt=false]{libertine}
\usepackage[scaled=0.83]{beramono}
\usepackage[libertine]{newtxmath}
\usepackage[usenames,dvipsnames]{xcolor}
\usepackage{graphicx}
\usepackage{bm}
\usepackage{array,multirow}
\usepackage{booktabs}
\usepackage{mathtools}
\usepackage{longtable}
\usepackage{pdflscape}
\mathtoolsset{showonlyrefs}

\usepackage{definitions}

\allowdisplaybreaks
\begin{document}
 
\title{Ultra High Energy Cosmic Rays from the Local Void}

\author[1]{Michael J. Padgett\thanks{michael.j.padgett@vanderbilt.edu}}
\author[1]{Thomas W. Kephart\thanks{tom.kephart@gmail.com}}
\affil[]{Department of Physics and Astronomy, Vanderbilt University, Nashville, TN 37235}

\date{\today}
\maketitle

\begin{abstract}
Ultra high energy cosmic rays have been see coming from the direction of the local cosmic void.
We use this fact to argue that  at least some of these these cosmic rays are relatively light magnetic monopoles and
 that their relative fraction  above 10$^{20}$ eV can be found from full sky observations.

\end{abstract}

\section{Introduction}
\label{section_introduction}
One of the highest energy cosmic ray (CR) ever seen,  the  Amaterasu  particle  found by the Telescope Array Group \cite{TelescopeArray:2023sbd}, arrived from the direction of the local cosmic void. The central value of its measured energy was $2.46\times 10^{20}$ eV.  As there is no apparent source of high energy protons or nuclei in the local void, and since the energy of such normal particles could not cross the void unless it started with much more energy than the observed Amaterasu energy, this was taken as evidence that the Amaterasu  particle must have been a magnetic monopole \cite{Frampton:2024shp}, because free magnetic monopoles can naturally have  energies  above $10^{20}$ eV \cite{Kephart:1995bi,Wick:2000yc}.

The reason protons  cannot make it across the void is that the Greisen–Zatsepin–Kuzmin \cite{Greisen:1966jv,Zatsepin:1966jv} limit of $E\sim 5\times 10^{19}$ eV (or GZK cutoff) is the upper bound on the   energy for a   cosmic ray protons traveling from their production in very distant galaxies through the intergalactic medium to us. The limit is caused by the cosmic ray proton energy being degraded by interactions with cosmic microwave background photons via the process $p+\gamma\rightarrow \Delta^* \rightarrow p+\pi^0$, which has a mean free path (mfp) of approximately  $ 6$ Mpc.  The recently observed Amaterasu cosmic ray particle  \cite{TelescopeArray:2023sbd} arrived from the direction of the local cosmic void, which subtends a solid angle of approximately $1.6\, \pi$ and is at least 45 Mpc across  \cite{TullyFisher,Tully:2019ngb}. It is highly unlikely that a trans GZK cosmic ray proton could cross such a distance. To do this the particle would need an initial energy of about $10^{21}$ eV on the other side of the void, which is a remote possibility, given known acceleration mechanisms. Figure 1 contains a Hillas plot which illustrates the criteria for acceleration of these CRs by relating the magnetic field and size of sources required to adequately contain the particles long enough to be accelerated \cite{Ptitsyna:2010Hillas, Bustamante:2009Acceleration}.

\begin{figure}[htbp]
\centering
      \includegraphics[width=0.8\textwidth,angle=0]{Hillas_Plot}
        \caption{Hillas plot for a proton (934 EeV) and Fe nucleus (240 EeV). Known acceleration sources are pictured}
  \label{***}
\end{figure}

That leads us to conclude that one of the following three possibilities is true; (i) there is a source of ultra high energy cosmic rays (UHECRs) within the void, (ii) the path of the  Amaterasu particle was bent, most likely within the Milky Way (see e.g., \cite{Anchordoqui:2018qom}), or (iii) the Amaterasu particle is  a magnetic monopole (MM) \cite{Kephart:1995bi,Wick:2000yc}. Free MMs need no  source, as they  have been wandering the Universe ever since their production in an early Universe gauge theory phase transition. They are also not subject to the GZK cutoff.

 Option (i), a UHECR source within the local void is possible  and has been analyzed in detail by \cite{Unger:2023hnu}. While there are a few dwarf galaxies in the local void, one expects the chances of producing UHECRs within the void is diminished by the ratio
of the density of galaxies in the local group compared to the density off galaxies in the local void, roughly a factor of $10^3$. This renders option (i) unlikely. Option (ii), could be a UHECR proton or nucleus that appears to have originated in the local void, but in actuality was produced elsewhere, say in the local group, and who's path was bent by interacting with a local object containing a strong 
extended B-field to make it appear as if the particle came from the local void. Again  this possibility is unlikely. A details  analysis of this type has been carried out by Anchordoqui as reported in his review   \cite{Anchordoqui:2018qom}. For other constraints on nearby sources of UHECRs see \cite{Kuznetsov:2023jfw}.
These facts led to option (iii),  the conclusion that the Amaterasu particle is a magnetic monopole \cite{Frampton:2024shp}.

If we assume a magnetic monopole  is the solution of the  Amaterasu particle mystery, then it must be relativistic, which means it is relatively light compared to the typical $\sim 10^{16}$ GeV GUT scale.  That puts a strong constraint on models where magnetic monopoles appear. However, there is still a large class of such models that can accommodate such monopoles as we will discuss below. 


The local void is a large irregularly shaped region and we live adjacent to it. It is difficult to be precise about ray tracing from our location back through the void, especially near its edges, so we will assume a UHECR coming from one radian of the center of the void coordinates is a ``void CR (void ray)'' and others are not. This allows us to calculate the surface density
(void rays per square solid angle) and compare it to the non-void ray density. That in turn will help us estimate the observed MM UHECR  fractional rate.

The rest of this paper includes: in section II a discussion of galactic and extragalactic magnetic fields and typical  MM energies; in section III a summary of observed UHECR and their trajectories for energies above the GZK cut-off; in IV an estimate of the MM fraction of UHECRs as a function of energy; and in section V  a discussion and our conclusions.

\section{Galactic and Extragalactic Magnetic Fields and Associated Magnetic Monopole Energy}

Let us begin by recalling an important relationship between magnetic fields and magnetic monopoles. Monopoles are accelerated along 
galactic magnetic fields lines and  consume those fields in the process. Requiring the fields are regenerated faster by galactic dynamos
than they are consumed  results in the Parker bound on the monopole flux
 ${\cal F} \le 10^{-15}$/$\rm{cm}^2$/s/sr. The Parker bound is several orders of magnitude
above the observed UHECR flux,
hence does not constrain MM UHECRs.

\begin{figure}[htbp]
\centering
      \includegraphics[width=0.7\textwidth,angle=0]{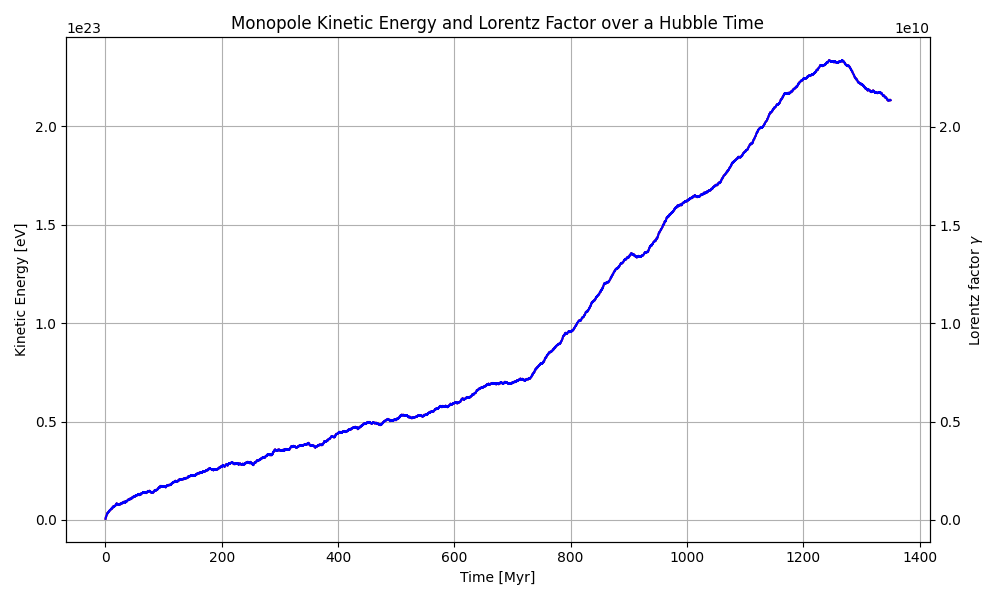}
        \caption{Average of random walks for 20 MMs in the local group with $m=10^4\,\mathrm{GeV}$}
  \label{RandomWalk}
\end{figure}

The kinetic energy gained by a magnetic monopole traveling along a magnetic
field of coherence length $\xi$ is \cite{Kephart:1995bi}
\begin{equation}
E
\sim\,g\,B\,\xi\, 
\label{Ekin}
\end{equation}
where
\begin{equation}
g=e/2\alpha=3.3\times10^{-8} \;\rm{esu} \;\;
({\rm or}\; 3.3\times 10^{-8} dynes/G)
\label{charge}
\end{equation}
is the magnetic charge according to the Dirac quantization condition, $\alpha$ is the fine structure constant and
$B$ is the magnetic field strength.
Since their production in the early Universe we expect that monopoles have random--walked through a large number $n$ of domains
of coherent fields and this would increase their resulting energy by roughly $\sqrt{n}$. However, the domains and magnetic fields  are all of different lengths.
Magnetic fields in galaxies, galaxy cluster, AGN jets, etc. range from 0.1--100 $\mu G$ while their coherence lengths range from $10^{-4}-30$ Mpc. This leads to monopole energies in the range 
$$1.7\times 10^{20}\,\, to\,\, 5\times 10^{23}\,\, \rm{eV}.$$
See \cite{Wick:2000yc} for more details, and for a more recent comprehensive study see \cite{Perri:2023ncd}.
As an example of expected MM energies
Figure \ref{RandomWalk} displays the $\gamma$ value and energy of a typical MM of mass $m=10^4\,\mathrm{GeV}$ as it random walks through our local extra galactic environment for a Hubble time, $t=\frac{1}{H_0}$.
The energy of the MM in Figure 2 reaches $E\sim2.4\times 10^{23}\, \rm{eV}$ with it's Lorentz factor reaching up to $\gamma\sim2.4\times 10^{10}$ showing that these MMs are highly relativistic which is required for the detection of UHECR primaries as will be discussed in Section 3.

It is of note that the energy value for the MM in Figure 2   marks a new source of UHECRs that can be expected to be a nearly isotropic spectrum. 

The current strongest bound on the flux of relativistic monopoles comes from Ice Cube and is \cite{IceCube:2021eye}
$${\cal F}_{_{MM}}\le 2\times 10^{-19} cm^{-2}s^{-1}sr^{-1},$$
which is well below the Parker bound. 
However,  the observed flux of UHECRs is only about one per $km^2$ per century, or
$${\cal F}_{_{UHECR}} \sim 3\times 10^{-20} cm^{-2}s^{-1}sr^{-1},$$
which is still below the Ice Cube bound.

\section{Observed UHECR and their Trajectories}


The highest initial energy UHECR ever seen is the Fly's Eye event \cite{HIRES:1994ijd} at energy $E_{FE} = 3.2 \times 10^{20}$ eV. The Amaterasu event and two events from AGASA \cite{Shinozaki:2006kk} at
nominal energies $E_{A1} = 2.4 \times 10^{20}$ eV and
$E_{A2} = 2.13 \times 10^{20}$ eV are all within 1 $\sigma$ of each other, hence within errors of being second in energy to
the Fly's Eye  event.  AGASA has reported a total eleven events with energy above $10^{20}$ eV, while
the Pierre Auger experiment has reported  20   events  between $1.10$ and $1.66 \times 10^{20}$ eV \cite{PierreAuger:2022qcg,PierreAuger:2022axr}.\\

Our interest is in the arrival directions of all these super GZK cosmic rays. Is the number suppressed from the direction of the local void? Little or no suppression would suggest a large magnetic monopoles component, while a strong suppression would suggest they are mostly or all protons and nuclei.

\vspace{0.5cm}
\noindent
{\it Monopole Direction:}

We first recall that the direction of the center of the local void is \cite{TullyFisher,Tully:2019ngb}

\begin{center}
(RA, Dec) =  $(279.5^{\circ},18.0^{\circ},)$
\end{center}

\noindent
while the arrival direction of the Amaterasu comic ray was 

\begin{center}
(RA, Dec) = $(255.9^{\circ}, 16.1^{\circ})$
\end{center}

\noindent
which is well within the direction of the local void. Assuming the Amaterasu particle was a proton or nucleus, then  if backtracked through the galactic field, due to its high energy, it still
appears to originate within the local void \cite{Unger:2023hnu}.
Given that there are no apparent sources of UHECR protons or nuclei within the local void and that it is highly unlikely that such a trajectory could have been bent by a foreground object within our galaxy, one is  led to conclude that the Amaterasu particle was a magnetic monopole  \cite{Frampton:2024shp}. \\

Appendix A contains the 36 most energetic events of the 720 trans-GZK UHECRs detected by PAO and AGASA as well as the OMG and Amaterasu particles. Skymaps can then be generated using this data to differentiate void- and non-void rays. Figure 3 is a skymap of all 720  UHECRs   where the void- and non-void rays are colored differently.

\begin{figure}[htbp]
\centering
      \includegraphics[width=0.8\textwidth,angle=0]{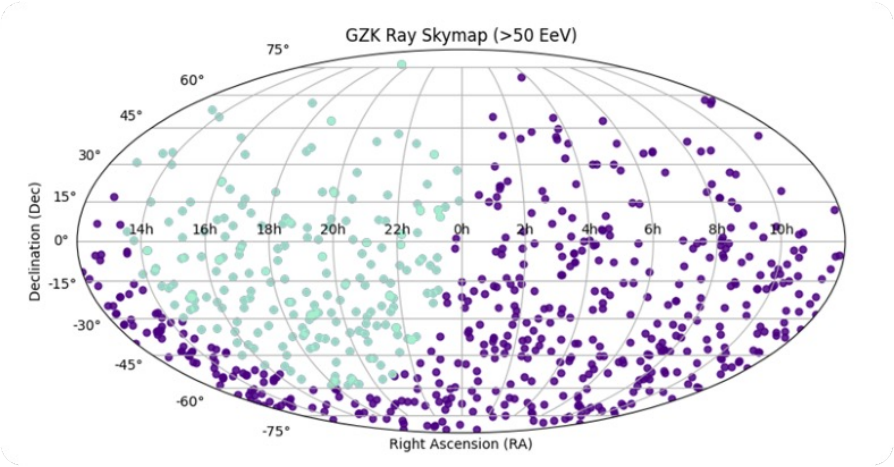}
        \caption{720 cosmic rays above the GZK limit from PAO and AGASA (including the Fly's Eye (~320 EeV) and Amaterasu particle). Void rays are colored in cyan and non-void rays in purple.}
  \label{***}
\end{figure}

We can then focus on only the void rays and break them down by their energy, shown in Figure 4, to get a more complete picture.

\begin{figure}[htbp]
\centering
      \includegraphics[width=0.8\textwidth,angle=0]{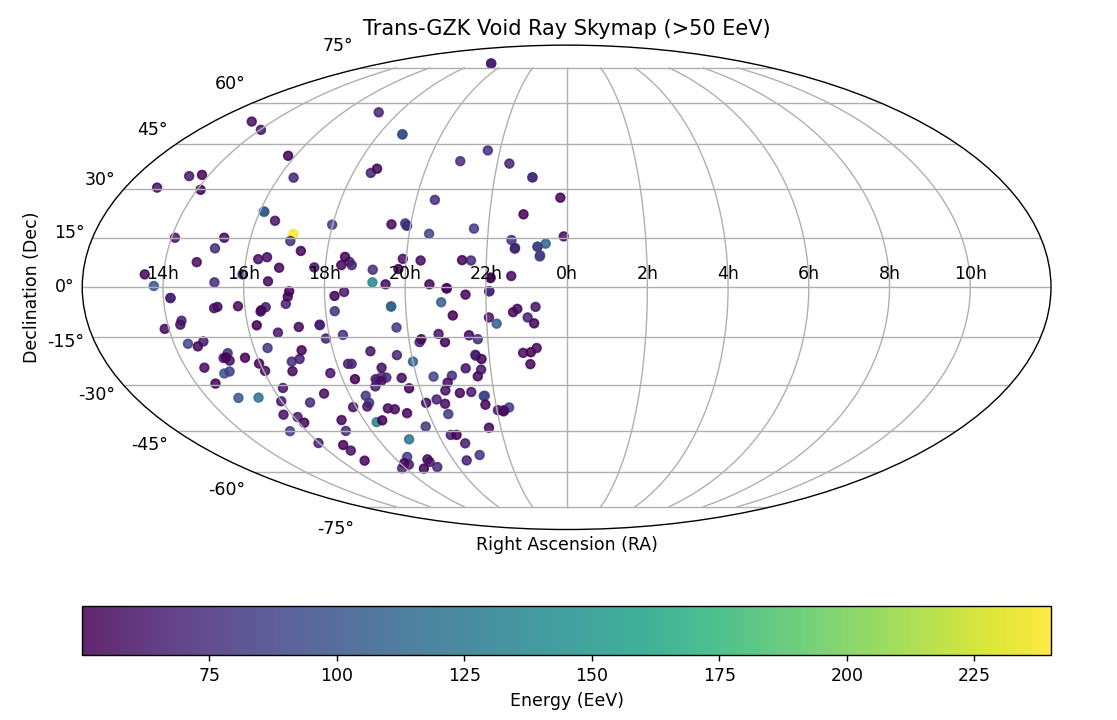}
        \caption{Trans-GZK void rays broken down by their reconstructed energy on arrival. The Amaterasu particle stands out as it is shown in yellow.}
  \label{***}
\end{figure}

\newpage

Since observed UHECR primaries are   highly relativistic this constrains the monopole mass $M$ to be
$$
M \le 10^{8} \,\, \rm{GeV}.
$$
Requiring the phase transition where monopoles are produced to be above the electroweak scale gives a  lower bound on the monopole mass 
$$
M  \ge \frac{M_{EW}}{\alpha}  \sim 10^{4} \,\, \rm{GeV}.
$$
This rules out  standard grand unification via the $SU(5), \, SO(10), E_6$ chain and suggests it is more likely to be a quiver gauge theory of the type discussed below.\\

\section{Estimate of the Magnetic Monopole Fraction of UHECR Flux}

By utilizing the trans-GZK UHECR data, we can weigh these values using Sommers relative exposure \cite{Sommers:2000relative} with the cited exposure values for these experiments as well as their geographic positions \cite{AGASA:exposure, PAO:exposure}. We can then generate a graph for the ratio of void- to non-void rays shown as the blue line in Figure 5.

This ratio is important because as MMs are produced during the phase transition in the early Universe, they do not require sources. This means that as the energy increases past the GZK cutoff, if the UHECR flux consists of MMs, we should observe a near isotropic intensity of UHECRs. The assumption we make in this statement is that no other acceleration mechanism turns on that would produce UHECRs with $E\sim10^{20}\,\mathrm{eV}$ within $\sim50\,\mathrm{Mpc}$. A study was done when the OMG particle was detected and found no sources within $\sim50\,\mathrm{Mpc}$, so this is likely to be a good assumption  \cite{Elbert:1994flyseye}. 

We can calculate a pre-GZK interval, $\Delta E_0$, that is used to calculate a baseline ratio

\begin{equation*}
    R_0=\frac{I_{v}(\Delta E_0)}{I_{nv}(\Delta E_0)}
\end{equation*}

For the following calculations we use $\Delta E_0=(30-50)\,\mathrm{EeV}$. We can then approximate the pseudo-intensity of MMs

\begin{equation*}
    I_{MM}(E)=I_v(E)-R_0I_{nv}(E)
\end{equation*}
and also clamp the MM pseudo-intensity to zero if the value fluctuates below the baseline ratio to avoid negative values. We will then approximate the MM fraction 

\begin{equation*}
    f_{MM}(E)=\frac{I_{MM}(E)}{I_{v}(E)}
\end{equation*}

We will next approximate the error on this fraction by propagating the Poissonian error evident in our original data bins

\begin{equation*}
    \sigma_{I_{MM}}^2=\sigma_{I_v}^2+R_0^2\sigma_{nv}^2+I_{nv}^2\sigma_{R_0}^2
\end{equation*}

Using our UHECR data we  produce a graph of the MM fraction shown as the orange line in Figure 5.

\begin{figure}[htbp]
\centering
      \includegraphics[width=0.8\textwidth,angle=0]{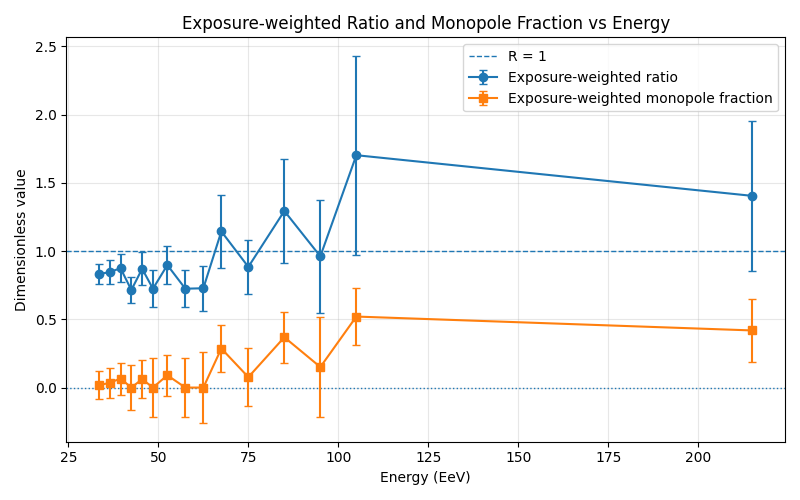}
        \caption{Void-Nonvoid ratio and MM fraction of UHECRs}
  \label{***}
\end{figure}

Finally we generate Table 1 for the MM fraction, $f_{MM}$, and the statistical significance given the associated uncertainties, $Z_{MM}=\frac{I_{MM}}{\sigma_{I_{MM}}}$.
The interpretation of the $Z_{MM}$ value is as follows

\begin{equation*}
Z_{MM}
\begin{cases}
    \sim1:\,\,\,\text{Statistically Insignificant}\\
    \sim2:\,\,\,\text{In-tension}\\
    \gtrsim3:\,\,\,\text{Evidence}
\end{cases}
\end{equation*}

\begin{table}[htbp]
\centering
\caption{MM fraction values as a function of energy, $f_{MM}$ with their statistical significance, $Z_{MM}$\\ 
\vspace{.4cm}\\}
\label{tab:f^{uw}_{MM}}
\renewcommand{\arraystretch}{1.3}
\begin{tabular}{|c|c|c|}
\hline
$\Delta E~\mathrm{[EeV]}$ & $f_{MM}$ & $Z_{MM}$\\
\hline
$(32-35)$ & 0.017 & 0.165\\
$(35-38)$ & 0.035 & 0.312\\
$(38-41)$   & 0.066 & 0.540\\
$(41-44)$   & 0 & -0.931\\
$(44-47)$   & 0.062 & 0.423\\
$(47-50)$   & 0 & -0.636\\
$(50-55)$   & 0.090 & 0.565\\
$(55-60)$   & 0 & -0.636\\
$(60-65)$   & 0 & -0.505\\
$(65-70)$   & 0.286 & 1.318\\
$(70-80)$   & 0.076 & 0.344\\
$(80-90)$   & 0.369 & 1.43\\
$(90-100)$   & 0.150 & 0.366\\
$(100-110)$   & 0.520 & 1.504\\
$(110-320)$   & 0.419 & 1.246\\
\hline
\end{tabular}
\end{table}

Even though the $Z_{MM}$ values contained in Table 1 don't show a smoking gun for a dominant MM fraction of observed UHECRs, they do contain a clear pattern of variation around zero below the GZK cutoff, of  being clearly positive above $\approx65~\mathrm{EeV}$ and even hinting towards being in tension with the (100-110) EeV range. As more UHECRs above the GZK cutoff are observed, we expect better statistics for the $Z_{MM}$ values with a trend to higher values indicating evidence of discovery. 

\vspace{0.5cm}
\noindent
\section{Models with Light Magnetic Monopoles} 

Quiver gauge theories (For a review see \cite{Frampton:2007fr}. Early papers on quiver gauge theories include
\cite{Kim:1980yk,Kachru:1998ys,Lawrence:1998ja,Frampton:2000mq}.) have separated gauge and flavor groups, and hence avoid proton decay for low scale spontaneous symmetry breaking (SSB).  If a $U(1)$ appears at this SSB scale, then there is an associated light  magnetic monopoles.  The simplest examples of this are the Pati-Salam model (PS) \cite{Pati:1973uk,Pati:1974yy} and the Trinification model (T) \cite{Trinification} with gauge groups $SU(4)\times SU(2)\times SU(2)$ and $SU(3)\times SU(3)\times SU(3)$ respectively. Monopoles in these theories have been studies in e.g., \cite{Kephart:2025tik}.

The appearance of monopoles in quiver gauge theories
which generalize the original PS and T models has
been analyzed in \cite{Kephart:2001ix,Kephart:2006zd,Kephart:2017esj,Sheridan:2022qku}. The last two of these papers used the
LieART platform \cite{Feger:2012bs,Feger:2019tvk}
to study extensions of PS and T, for
gauge groups with three factors $G =SU(a) \times SU(b) \times SU(c)$ with $a \geq 3, b, c \geq 2$ subject to
an upper limit $\dim[G]= (a^2 + b^2 + c^2 -3) \leq 78 =\dim[E(6)]$.
By restricting the size of $G$, a manageably finite, but large, number
of models were selected and studied more completely  than was previously accomplished by hand.

Initially in quiver gauge theories, the chiral fermions are all put into bifundamental representations,
then permissible additional anomaly-free non-bifundamental combinations
are added. For example, in an $(abc) = (333)$ theory, one can add
$3(3,1,1)+3(1,{\bar 3},1)+({\bar 3},3,1)$ while keeping anomaly cancellation,
although this theory is no longer a subgroup of an $E_6$ grand unified theory (GUT).

Once we depart from the framework of GUTS, leptons with fractional
electric charges {\it e.g.} $(\pm \frac{e}{2}, \pm \frac{e}{6})$,
appear. Thus, if the Amaterasu
cosmic ray is a light monopole, it suggest the probable existence of fractionally-charged
leptons in particle theory \cite{Kephart:2001ix,Kephart:2006zd,Kephart:2017esj,Sheridan:2022qku}. Fractional electric charged particles
in turn predicts multiply charged monopoles in order that the Dirac
relation $ge=n\hbar/2$ is maintained.

The mass of the magnetic monopoles can naturally be at an intermediate
scale  and new particle physics, 
like fractional charged lepton,  should be expected
at a few $TeV$ scale. Indirect evidence of monopoles may be accessible to the LHC 
where there are now dedicated monopole searches such as the MoEDAL 
experiment \cite{MoEDAL:2009jwa,MoEDAL:2014ttp}.
Searches for the additional light particles, such as fractionally-charged
leptons and non standard charge quarks as predicted by such magnetic monopole theories, can be performed
at the Large Hadron Collider (LHC). 

Since we want a high energy monopole to look like the Amaterasu event, our interest is in baryonic monopoles, bound states of monopoles with both color magnetic and $U(1)_{EM}$ magnetic charge. These particles exist in various quiver gauge theories \cite{Kephart:2025tik}, but for
the present discussion we focus on a model independent analysis. For a hard collision involving one of the constituent quarks in normal baryon, the quark winds up on the end of a QCD electric flux tube that typically breaks once it is long enough to fragment into mesons.
If the initial collision is between a proton and a nucleus in our upper atmosphere, then a cascade follows and the remnants air shower can be seen in a detector on the surface of the earth.
For a collision involving an atmospheric nucleon and a individual monopole within the baryonic monopoles, a string (color magnetic flux tube) forms, but it can not break unless there is enough energy
to produce a monopole-antimonople pair. Consequently in most such collisions the color magnetic string oscillates and radiates light particles. But while this string is still stretched, the 
excited baryonic monopole's cross section is dramatically increased and further scattering
takes place that can mimic the air shower of a ultra high energy proton or nucleus \cite{Wick:2000yc}.

\vspace{0.5cm}
\noindent
\section{Discussion and Conclusion}
The Local Void is not the only void in our vicinity. But other neighborhood voids like the Hercules Void and the Sculptor Void both are more likely to have intervening foreground sources of UHE protons and/or nuclei, so we have focused our study of UHECRs on the Local Void.

Cosmic rays have historically played a major role in particle physics, and include the original discoveries of the positron and the muon. The Amaterasu cosmic ray is only one event but one of the most energetic
cosmic rays ever recorded and   pointing
back to the local void where there is no obvious source. It has led us to consider all other UHECRs  and conclude that a substantial fraction can be coming from the local void.
This renders it unlikely that the primary is a proton or nucleus and leads to our favored interpretation that
magnetic monopoles may have finally been discovered as predicted in the elegant  theory of
Dirac in 1931. (That the Amaterasu particle may be a magnetic monopole has also recently been suggested by \cite{Cho:2023krz}.) If this is correct, it  has implications for what is the
most likely extension of the standard model of particle physics.
We have suggested that an obvious choice is a  quiver gauge theory with chiral fermions in bifundamental representations
for which fractionally charged leptons or other associated states  could be discovered at the LHC.
Once again  remarkable and unexpected observation in cosmic rays, may, provide a potentially
extraordinary step forward in the theory of particle physics.

\newpage
\begin{landscape}
\section{Appendix A}
\scriptsize
\setlength{\tabcolsep}{2pt}
\renewcommand{\arraystretch}{1.1}

\begin{longtable}{|l|l|l|l|l|l|l|l|l|}
\hline
\caption{Trans-GZK UHECRs}
\label{tab:uhecr}\\
\hline
        Cosmic Ray Experiment & Energy (EeV) & RA (hours) & RA (minutes) & RA (seconds) & RA (degrees) & Dec (degrees) & Galactic Longitude (degrees) & Galactic Latitude (degrees) \\
\hline
\endfirsthead

\hline
\multicolumn{9}{|c|}{\tablename\ \thetable\ -- continued from previous page} \\
\hline
\endhead

\hline
\multicolumn{9}{|r|}{Continued on next page} \\
\hline
\endfoot

\hline
\endlastfoot
        FE & 320 & ~ & ~ & ~ & 85.2 & 48 & 163.4 & 9.6 \\ \hline
        TA & 240 & ~ & ~ & ~ & 255.9 & 16.1 & ~ & ~ \\ \hline
        AGASA & 213 & 1 & 15 & 0 & 18.75 & 21.1 & 130.5 & -41.4 \\ \hline
        PAO & 165.5 & ~ & ~ & ~ & 128.9 & -52 & ~ & ~ \\ \hline
        PAO & 164.7 & ~ & ~ & ~ & 192.9 & -21.2 & ~ & ~ \\ \hline
        PAO & 155.2 & ~ & ~ & ~ & 107.2 & -47.6 & ~ & ~ \\ \hline
        PAO & 155.1 & ~ & ~ & ~ & 102.9 & -37.8 & ~ & ~ \\ \hline
        AGASA & 150 & 19 & 38 & 0 & 294.5 & -5.8 & 33.1 & -13.1 \\ \hline
        AGASA & 149.1 & ~ & ~ & ~ & ~ & ~ & 130.5 & -41.4 \\ \hline
        PAO & 148 & ~ & ~ & ~ & ~ & ~ & -57.2 & 41.8 \\ \hline
        PAO & 146.5 & ~ & ~ & ~ & 125 & -0.6 & ~ & ~ \\ \hline
        AGASA & 144 & 16 & 6 & 0 & 241.5 & 23 & 38.9 & 45.8 \\ \hline
        PAO & 139.9 & ~ & ~ & ~ & 287.8 & 1.5 & ~ & ~ \\ \hline
        AGASA & 134 & 18 & 45 & 0 & 281.25 & 48.3 & 77.6 & 20.9 \\ \hline
        PAO & 132.6 & ~ & ~ & ~ & 275 & -42.1 & ~ & ~ \\ \hline
        PAO & 131.6 & ~ & ~ & ~ & 107.8 & -44.7 & ~ & ~ \\ \hline
        PAO & 130.6 & ~ & ~ & ~ & 340.6 & 12 & ~ & ~ \\ \hline
        PAO & 127 & ~ & ~ & ~ & 45.8 & -1.7 & ~ & ~ \\ \hline
        PAO & 124.9 & ~ & ~ & ~ & 284.8 & -48 & ~ & ~ \\ \hline
        PAO & 123.7 & ~ & ~ & ~ & 352.5 & -20.8 & ~ & ~ \\ \hline
        PAO & 122.3 & ~ & ~ & ~ & 175.6 & -37.7 & ~ & ~ \\ \hline
        PAO & 121.8 & ~ & ~ & ~ & 231.4 & -34 & ~ & ~ \\ \hline
        AGASA & 120 & 23 & 16 & 0 & 349 & 12.3 & 89.5 & -44.3 \\ \hline
        PAO & 115.8 & ~ & ~ & ~ & 150 & -10.4 & ~ & ~ \\ \hline
        PAO & 115.2 & ~ & ~ & ~ & 151.5 & -12.6 & ~ & ~ \\ \hline
        PAO & 113.3 & ~ & ~ & ~ & 21.7 & -13.8 & ~ & ~ \\ \hline
        PAO & 110.6 & ~ & ~ & ~ & 352.1 & 13.2 & ~ & ~ \\ \hline
        PAO & 109.7 & ~ & ~ & ~ & 133.6 & -38.3 & ~ & ~ \\ \hline
        PAO & 109.6 & ~ & ~ & ~ & 300 & -22.6 & ~ & ~ \\ \hline
        PAO & 107.8 & ~ & ~ & ~ & 344.1 & -71.5 & ~ & ~ \\ \hline
        PAO & 107.6 & ~ & ~ & ~ & 185.9 & -70.7 & ~ & ~ \\ \hline
        PAO & 107.1 & ~ & ~ & ~ & 83.1 & -51.1 & ~ & ~ \\ \hline
        PAO & 107 & ~ & ~ & ~ & 65.8 & -26.3 & ~ & ~ \\ \hline
        PAO & 106.8 & ~ & ~ & ~ & 313.3 & -4.5 & ~ & ~ \\ \hline
        PAO & 106.2 & ~ & ~ & ~ & 333.7 & -11 & ~ & ~ \\ \hline
        PAO & 105.7 & ~ & ~ & ~ & 164.6 & -20.1 & ~ & ~ \\ \hline
\end{longtable}
\end{landscape}

\end{document}